\documentclass[12pt,preprint]{aastex}

\slugcomment{Accepted: August 15, 2012}

\shorttitle{The Hubble Constant}
\shortauthors{Freedman et al.}

\begin{document}


\title{Carnegie Hubble Program: \\ A Mid-Infrared Calibration \\of the \\ Hubble Constant \\ }

\medskip
\medskip
\medskip
\medskip
\medskip

\author{\bf }
\author{\bf }
\author{\bf Wendy L. Freedman, Barry F. Madore}

\author{\bf   Victoria Scowcroft, Chris Burns,  Andy Monson }

\author{\bf   S. Eric Persson, Mark Seibert}

\affil{Observatories of the Carnegie
  Institution of Washington \\ 813 Santa Barbara St., Pasadena, CA
  ~~91101} 

\author{\bf Jane Rigby}
\affil{Observational Cosmology Lab, NASA Goddard Space Flight Center \\  Greenbelt MD 20771}

\author{\bf }
\author{\bf }
\author{\bf }
\author{\bf }
\author{\bf }

\medskip
\medskip
\medskip
\medskip

\email{wendy@obs.carnegiescience.edu,\\ barry@obs.carnegiescience.edu,\\ vs@obs.carnegiescience.edu,
 \\ cburns@obs.carnegiescience.edu, \\ amonson@obs.carnegiescience.edu,
  \\ persson@obs.carnegiescience.edu,  mseibert@obs.carnegiescience.edu, \\ Jane.R.Rigby@nasa.gov}



\vfill\eject

\begin{abstract}
Using a mid-infrared calibration of the Cepheid distance scale based
on recent observations  at 3.6~$\mu$m with the {\it Spitzer
  Space Telescope}, we have obtained  a new, high-accuracy calibration
of the Hubble constant. We have established the mid-IR zero point of
the Leavitt Law (the Cepheid Period-Luminosity relation) using
time-averaged 3.6~$\mu$m data for ten high-metallicity, Milky Way
Cepheids having independently-measured trigonometric parallaxes.  We
have adopted the slope of the PL relation using time-averaged
3.6~$\mu$m data for 80 long-period Large Magellanic Cloud (LMC)
Cepheids falling in the period range 0.8 $<$ log(P) $<$ 1.8.  We find
a new reddening-corrected distance to the LMC of 18.477 $\pm$ 0.033 (systematic)
mag. We re-examine the systematic uncertainties in $H_0$, also taking
into account new data over the past decade. In combination with the
new {\it Spitzer} calibration, the systematic uncertainty in $H_0$
over that obtained by the {\it Hubble Space Telescope} ({\it HST})
{\it Key Project} has decreased by over a factor of three. Applying
the {\it Spitzer} calibration to the {\it Key Project} sample, we find
a value of $H_0$ = 74.3 with a systematic uncertainty of $\pm$ 2.1
(systematic)~km s$^{-1}$ Mpc$^{-1}$, corresponding to a 2.8\%
systematic uncertainty in the Hubble constant. This result, in
combination with WMAP7 measurements of the cosmic microwave background
anisotropies and assuming a flat universe, yields a value of the
equation of state for dark energy, $w_0$ = -1.09 $\pm$
0.10. Alternatively, relaxing the constraints on flatness and the
numbers of relativistic species, and combining our results with those
of WMAP7, Type Ia supernovae and baryon acoustic oscillations yields
$w_0$ = -1.08 $\pm$ 0.10 and a value of N$_{eff}$ = 4.13 $\pm$ 0.67,
mildly consistent with the existence of a fourth neutrino species. 

\end{abstract}

.

\medskip
\medskip
\medskip
\medskip
\medskip
\medskip
\medskip
\medskip
\medskip
\vfill\eject
\section{Introduction}

Over the past several decades, there has been a steady increase in the
accuracy with which extragalactic distances and the Hubble constant
can be measured ({\it e.g.,} Freedman {\it et al}. 2001, hereafter F01; Riess {\it et
  al}. 2011; Komatsu {\it et al}. 2011; and for a review see Freedman
\& Madore 2010). This has resulted from a number of factors: the
availability, on the ground and especially in space, of high
throughput, high dynamic range optical CCDs and infrared arrays; the
multi-wavelength sensitivity of these devices; making it possible to
correct for systematic effects of reddening and metallicity; and to use a
wider range of methods for measuring relative distances beyond the
immediate reach of Cepheid variables. Just over a decade ago, there
was still debate over the value of the Hubble constant at a level of a
factor of two; today, there is the promise of measuring the Hubble
constant to an accuracy better than two percent.

In combination with other constraints ({\it e.g.,} the angular power
spectrum of cosmic microwave background anisotropies), an independent
measurement of H$_0$ to accuracy of better than a few percent can
provide critical constraints on the dark energy equation of state, the
spatial curvature of the universe, neutrino physics and general
relativity (see Suyu {\it et al}. 2012 for a recent discussion).  In
practice, an accurate value of H$_0$ provides a means of breaking the
degeneracies amongst several cosmological parameters. For example,
measurements of cosmic microwave background anisotropies yield
well-determined values of the products of $\Omega_m h^2$, $\Omega_b
h^2$ (where $\Omega_m$ and $\Omega_b$ are the matter and baryon
densities, respectively), but not the densities, nor H$_0$
independently.  There are other degeneracies between H$_0$ and the
equation of state, w$_0$, as well as its evolution, w$_a$; between
H$_0$ and the number of relativisitic neutrinos, N$_{eff}$ and the sum
of the masses of neutrinos; and between H$_0$ and $\sigma_8$, the
fluctuation of matter on 8 Mpc scales (for recent discussions, see
Dunkley {\it et al}. 2009; Komatsu {\it et al}. 2009). Hence, the
motivation for measuring an independent value of H$_0$ accurate to a
few percent has continued to increase.

As described in Freedman {\it et al}. (2011, hereafter F11), we have
begun a new Carnegie Hubble Program (CHP), specifically designed to
minimize and/or eliminate the remaining known systematics in the
measurement of the Hubble constant using mid-infrared data from NASA's
{\it Spitzer Space Telescope}. Here we report on a newly derived value
of the Hubble constant and its uncertainty, based on the first data
acquired from this program for Cepheids in the Milky Way and the
LMC. Our focus in this paper is primarily the zero point of the
Cepheid extragalactic distance scale, and a reassessment of the
systematic error budget.

\section{A New Mid-Infrared Zero-Point Calibration of the Hubble Constant}
\label{sec:newcal}

As discussed at length in F11, there are many advantages in acquiring
3.6 $\mu$m data compared to optical observations. The effects of
reddening are decreased ({\it e.g.}, A$_V$ / A(3.6 $\mu$m) $\sim$ 15,
and A(I) / A(3.6 $\mu$m) $\sim$ 8; see F11 and references therein).
Metallicity effects are both theoretically predicted and empirically
demonstrated to be smaller. In addition, the dispersion in the Leavitt
Law is known to be more than a factor of two smaller in the
mid-infrared than in the V-band. The CHP is designed to establish the
calibration of the Cepheid extragalactic distance scale at
mid-infrared wavelengths by observing galaxies with known Cepheids in
the Local Group and beyond; undertaking empirical tests for
metallicity effects; providing a mid-infrared calibration of the
Tully-Fisher relation and for the Type Ia supernova distance
scale. The ultimate goal is a measurement of the Hubble constant to
$\pm$2\% (statistical plus systematic) uncertainty.

\subsection{The CHP Data}

To date, as part of the CHP, we have obtained 3.6 and 4.5~$\mu$m
observations for a sample of 37 Galactic Cepheids (Monson {\it et al}.
2012), 10 of which have direct trigonometric parallaxes measured by
the {\it HST} Fine Guidance Sensors (Benedict {\it
  et al}.  2007). Twenty-four observations were made at each
wavelength for each Cepheid. Since the periods of these Cepheids are
known {\it a priori}, we were able to schedule these 24 {\it Spitzer}
observations with a roughly constant or uniform spacing over time for
each of the stars. The Milky Way Cepheids range in period from roughly
4 to 36 days; five of these Cepheids have periods greater than six
days. We have also obtained similarly high-quality, uniformly-sampled
data for 80 LMC Cepheids (Scowcroft ~{\it et al}. 2011, 2012), with periods
in the range of 6 to 60 days, distributed across the face of the
galaxy, and chosen to be relatively uncrowded, based on H-band images
from Persson {\it et al}. (2004).

Although not originally designed for a cosmic distance scale program,
{\it Spitzer} (Werner {\it et al}. 2004) has proven itself to be an
excellent combination of telescope and instruments to measure the
Cepheid distance scale at long wavelengths. The mid-IR Infrared Array
Camera (IRAC; Fazio {\it et al}. 2004) has the dynamic range,
sensitivity and the spatial resolution to be able to measure both the
brightest of the Milky Way (calibrator) Cepheids and (with the same
telescope, instrument and filters) target and measure Cepheids in the
LMC, as well as other nearby galaxies.

\subsection{The Milky Way and LMC Leavitt (Cepheid Period-Luminosity) Relations}

Historically, the overlap in period between the Milky Way and LMC
Cepheid calibrators has been small.  Beyond the LMC, extragalactic
distance scale measurements are necessarily limited to the brightest
and longest-period Cepheids (generally P $>$ 10 days.)  The Galactic
{\it HST} parallax sample contains only one truly long-period Cepheid,
$\ell$~Car at P = 35.5~days. We therefore use the larger Cepheid
sample in the LMC to define the slope and width of the long-period (P
$\ge$ 6 days) end of the Leavitt Law.  In Figure 1, we show the
extinction-corrected Leavitt relations for the sample of 80 LMC stars
with {\it Spitzer} data from Scowcroft {\it et al}. (2011, 2012). The
10 Milky Way calibrators with parallax measurements from Benedict {\it
  et al}. (2007) and new {\it Spitzer} observations from Monson {\it
  et al}. (2012)\footnote{We note that the photometry for two stars in
  Table 4 of Monson {\it et al}. has been updated by Benedict {\it et
    al.} private communication.)} are individually labelled. LMC data
for stars with P $<$ 6 days from Meixner {\it et al}. (2006) are also
plotted, but not included in the fits. We note that the slopes of the
mid-infrared Leavitt relations for both the Milky Way and the LMC are
consistent over the entire period range from 4 to 60 days.

Both the slope and dispersion of the Galactic sample alone are, to
within the measurement uncertainties, in agreement with the LMC
Cepheid Leavitt relation. An unweighted least-squares fit to the 10
Milky Way stars from Monson {\it et al}. (2012) gives a slope of -3.40
$\pm$ 0.11 , which is statistically in agreement with the more robust
value (-3.31 $\pm$ 0.05) determined from the 80 stars covering the
same period range in our LMC sample (Scowcroft et al. 2011,
2012). Table 8 of Monson et al. provides zero-points and slopes for
the 10 Milky Way Cepheids at 3.6 $\mu$m for a number of different
weighting schemes, as well as a zero point obtained by fixing and
adopting the LMC slope. The zero points are -5.81 (unweighted), and
-5.80 (for two different weighting schemes). For
comparison, fixing the slope to be that for the LMC, yields an
intercept of -5.80 (unweighted) or -5.79 (weighted).  In all cases, the computed uncertainty is 0.03-0.04,
and the agreement is excellent.

 The dispersion in the extinction-corrected Leavitt relation for the
 LMC at 3.6~$\mu$m amounts to only $\pm$0.106 mag, giving an
 uncertainty of $\pm$5\% in distance for a single Cepheid; the
 dispersion for the sample of ten Milky Way Cepheids is $\pm$0.104
 mag. Correcting for the tilt of the LMC ({\it e.g.}, Scowcroft et
 al. 2012), the scatter in the 3.6 ~$\mu$m Leavitt relation reduces
 from $\pm$0.106 to $\pm$0.100 ~mag. With this sample of Cepheids
 observed at these long wavelengths, the random error on the distance
 modulus to the LMC has been reduced to $\pm$0.100/ $\sqrt{80}$ =
   $\pm$0.011 mag.  We take the systematic error on the distance
   modulus to the LMC to be defined by the uncertainty in the Milky
   Way best-fit intercept, when both the slope and intercept are fit
   simultaneously for the 10 Milky Way Cepheids, given by $\pm$0.104 /
   $\sqrt{10}$ = 0.033 mag. Adopting the ten Galactic calibrators
   these data yield a true distance modulus to the LMC of 18.477 $\pm$
   0.011 ({\it statistical}) $\pm$ 0.033~mag ({\it systematic}).

\subsection{Applying the New {\it Spitzer} Calibration}

Most recent extragalactic studies have used the distance to the LMC as
fiducial. For example, the methodology of the {\it Key Project}
(F01) was to adopt an LMC true distance modulus
of 18.50 $\pm$ 0.10~mag, and a reddening to the LMC of E(B-V) = 0.10
mag.  Relative distance moduli of galaxies beyond the LMC were then
determined using the reddening-free Wesenheit function, $W = V- R
\times (V-I)$, where $R$ is the ratio of total to selective absorption.
The advantage of this approach is that given an updated zero point,
any offset can simply be applied to the entire {\it Key Project}
distance scale.

We apply our new {\it Spitzer} zero-point calibration to the Benedict
{\it et al}. (2007) Milky Way parallax stars, and combine it with the
{\it HST} {\it Key Project} data from F01.  We note that the mid-IR
{\it Spitzer}-based distance modulus for the LMC is 0.023~mag ($\sim$1.2\%
in distance) smaller than the value of 18.50~mag adopted by F01 as
part of the {\it Key Project}, thus increasing $H_0$ by 1.2\%.  In
addition, (as described in \S \ref{sec:metallicity}), switching from a
LMC-based zero point to the Milky Way calibration, the metallicity
correction to the {\it Key Project} sample is now also significantly
reduced (by 0.04 mag giving an additional 2\% increase in $H_0$). We
obtain a value of $H_0$ = 74.3 $\pm$ 2.1 ({\it systematic}) ~km
s$^{-1}$ Mpc$^{-1}$.  This value of the Hubble constant is in
excellent agreement with that of F01, as well as more recent
determinations by Riess {\it et al}.  (2011), and Komatsu {\it et al}.
(2011). We provide a detailed discussion of the systematic uncertainty
on this value in the following section.

\section{Decreasing / Eliminating the Systematics}
\label{sec:systematics}

At the conclusion of the {\it Key Project}, F01 quantified the
outstanding sources of {\it systematic} uncertainty (their Table 14)
known to be affecting the value of the Hubble constant at that time.
These included: the absolute zero point of the Leavitt Law, which was
explicitly tied to the distance to the LMC ($\pm$5\%); the uncertainty
in the metallicity correction to the Leavitt period-luminosity (PL)
relation, resulting from the systematic offset in mean metallicity
between the LMC and many of the more distant (higher metallicity)
spiral galaxies ($\pm$4\%); the cumulative uncertainties resulting
from the cross-calibrating of instruments, filters and detectors from
the ground to space ($\pm$3.5\%); systematic reddening errors
($\pm$1\%); bias in fitting the PL relation to truncated data
($\pm$1\%); crowding or blending of images based on artificial star
tests ($0,+$5\%); and finally allowing for bulk flows on large scales
($\pm$5\%).

The mid-infrared / trigonometric parallax calibration for Galactic
Cepheids immediately solves two outstanding problems. It provides an
accurate (geometric) foundation upon which to set the zero point of
the Leavitt Law, at any given wavelength. In addition, it provides
high-metallicity Cepheid calibrators that are more comparable to the
bulk of high-metallity Cepheids in the target galaxies used for the
{\it HST} {\it Key Project}, as well as other determinations of $H_0$. 

Below we discuss the two systematic errors for which there is
significant quantitative improvement resulting from the new {\it Spitzer} data. We
then discuss the status of other terms entering the systematic error
budget.

\subsection{Absolute Zero Point of the PL Relation}
\label{sec:absolute}

As we saw in \S \ref{sec:newcal}, the measured scatter in the Milky Way
Leavitt relation of $\pm$0.104 mag suggests that the ten Galactic
calibrators define the zero point to $\pm$0.104/$\sqrt{10}$ = 0.033
mag. We adopt this value as the systematic error on the zero point of
the Cepheid PL relation.  Despite the small sample of Galactic
calibrators, the systematic error on the zero point of the Cepheid
Leavitt relation is already reduced to only 1.7$\%$ in distance; i.e.,
a factor of three better than the quoted uncertainty of the {\it HST}
{\it Key Project} zero point.  This uncertainty is also a measure of
the {\it systematic error} on the distance to the LMC.

We have been awarded further {\it Spitzer} time to observe a
sample of nearby Galactic Cepheids, so that ultimately GAIA can
provide accurate parallaxes for a sample size comparable to that of
the LMC. Windmark {\it et al}. (2011) have recently discussed the
potential of using GAIA to determine the Cepheid zero point. Although
they estimate that approximately 9000 Cepheids will be within reach of
GAIA, the overall accuracy will be limited by systematic effects,
primarily reddening. Hence obtaining mid-infrared observations for
these Milky Way Cepheids remains critical.

\subsection{Metallicity Dependence and Offsets}
\label{sec:metallicity}

A number of observations and tests are built into the CHP {\it
  Spitzer} program (see F11) to quantify the
magnitude of any residual metallicity effect. The first of these tests
is discussed in F11 where the deviations of
individual 3.6~$\mu$m LMC Cepheid magnitudes from the PL relation, as
a function of spectroscopic [Fe/H] metal abundances (from Romaniello
{\it et al}. 2008), exhibit only a very shallow (and statistically
insignificant) slope of -0.09 $\pm$ 0.29 mag/dex. Moreover, Figure 7
of F11 shows no evidence of any metallicity effect
for three Galactic Cepheids for which there are both 3.6~$\mu$m data
and [Fe/H] measurements. As we noted above, in moving to a zero-point
calibration of the Milky Way Cepheid PL relation at 3.6~$\mu$m, the
Cepheid zero-point calibration can now be directly established at high
metallicity, avoiding the intermediate step of calibrating with the
(lower-metallicity) LMC.

The metallicities (12 + log (O/H)) of the LMC, Milky Way and the mean
of the {\it Key Project} spiral galaxies are 8.50, 8.70, and 8.84 dex,
respectively (F01 and references therein; Allende-Prieto {\it et al}.
2001). For the {\it Key Project}, a slope of the metallicity
correction of -0.20 $\pm$ 0.2 mag/dex was adopted, where one dex
corresponds to a factor of ten in metallicity. Specifically, a
correction to the distance modulus of $\Delta \mu$ = -0.20 $\times$
([O/H] - [O/H]$_{LMC}$) was made.  F01 adopted a metallicity
correction of -0.20 $\times$ 0.34 = -0.068 mag, corresponding to a
decrease in the Hubble constant of 3.5\%, and adopted an uncertainty
corresponding to the entire correction. Now, correcting back from the
LMC to solar metallicity (0.20 mag/dex $\times$ 0.20 dex = 0.040 mag) results
in a differential correction to the Hubble constant of +2\%. In the
past we conservatively adopted the total magnitude of the correction
as being equivalent to its own systematic uncertainty.  The difference
between the {\it Key Project} sample and the Milky Way now implies a
correction of 0.2 mag/dex $\times$ 0.14 dex = 0.028 mag, or 1.4\% in
the distance scale.

In Figure 2 we show updated and revised plots of magnitude and color  residuals
from the mid-IR PL relations plotted as a function of spectroscopic
atmospheric [Fe/H] metallicities from Romaniello {\it et al}. (2008).
This plot supersedes earlier versions given in Freedman \& Madore
(2010) and Freedman {\it et al}. (2011) as it now uses the final
magnitudes and PL fits for the entire  LMC sample given in Scowcroft{\it et
  al}. (2011, 2012) and extends the Milky Way sample to significantly
higher metallicities using the newly published Galactic Cepheid mid-IR
data of Monson {\it et al}. (2012).  The formal solutions are $\Delta
[3.6] = +0.07 (\pm0.18)$ [Fe/H] $ - 0.01 (\pm0.06)$ and $\Delta [4.5] =
+0.04 (\pm0.19) [Fe/H] + 0.04 (\pm0.06)$. These data are consistent
with no significant correlation between the metallicity and the 3.6
or 4.5 $\mu$m Cepheid magnitudes over the metallicity range $-0.6 <$
[Fe/H] $< +0.2$ .

\subsection{Other Systematic Effects}

\noindent {\it WFPC2 Zero Point / Instrumental Systematics:} As
discussed in Stetson (1998), the Wide Field and Planetary Camera 2
(WFPC2) on {HST} had an imperfect charge transfer efficiency. To
quantify the uncertainties, Stetson carried out an extensive
comparison of WFPC2 and ground-based photometry for Milky Way globular
cluster stars. He found that the formal standard errors in this
comparison were significantly less than 1\%, but concluded that the
true external uncertainties were more likely at least of order 1\%. A
difference of 0.07 $\pm$ 0.02 mag was found between the ground-based
and WFPC2 photometry, and this correction was applied to the latter in
F01.  We note that in Table 14 of F01, the magnitude of this offset
(0.07 mag or 3.5\%) rather than the uncertainty (0.02 mag or 1\%) was
tabulated. In this paper, we quote the original uncertainty
consistent with that determined by Stetson (1998).

\noindent {\it Reddening:} From the optical through the infrared, the
effect of interstellar extinction is a generally declining function of
increasing wavelength.  A significant advantage of observing Cepheids
in the mid-infrared is therefore a reduced sensitivity to extinction
(both Galactic and extragalactic). The interstellar extinction law at
mid-infrared wavelengths has now been measured by a number of authors
(see F11 and references therein).  The shape of
the extinction curve shows some variation between different
sightlines, with an observed range of $A_{\lambda}/A_V$ = 0.058 to
0.071 at 3.6 $\mu$m, and 0.023 to 0.060 at 4.5 $\mu$m.  However, the
extinction measured in magnitudes in the mid-IR, as compared to
optical $V$-band data, for example, is reduced by factors of 14-17 at
3.6 $\mu$m, and 16-43 at 4.5 $\mu$m.  Thus, moving to the mid-IR
reduces, by more than an order of magnitude, the sensitivity of the
zero point of the Cepheid distance scale to reddening
corrections. Although uncertainties in reddening may contribute to
statistical uncertainties (at the $<$1\% level) for individual
galaxies, as discussed in F01), they are no
longer a significant contributor to the overall systematic error budget.

\noindent {\it PL Fitting Bias:} Apparent magnitude cut-offs in the
discovery and measurement of Cepheids at the detection limits of
surveys can, in principle, give rise to biased fitting errors. This
effect is a decreasing function of width (i.e., intrinsic dispersion)
of the PL relation, and hence, is again ameliorated by working at
mid-infrared wavelengths, or using the reddening-free magnitude, W.
By performing successive period cuts, F01
determined that the bias was negligible for the {\it Key Project}
sample, with the mean correction for the sample amounting to 0.01
mag. A period cut was applied to the shortest period Cepheids to
correct for this small bias.

\noindent {\it Crowding:} At present, crowding is not an issue for the Milky Way
Cepheids, which are bright and isolated. It is also not a significant
issue for the LMC sample of Scowcroft {\it et al}.  (2011, 2012), where the
LMC sample Cepheids were pre-selected on the basis of near-infrared
images to be isolated. In the case of the more distant {\it HST} {\it Key Project} sample, a
published uncertainty of $<$0.02 mag (``even in the most problematic
cases'') was given by Ferrarese et al. (2000) based on artificial star
tests. We note that F01 adopted an
uncertainty of 5\% (0.10 mag) due to crowding; however, no
justification was provided for this larger adopted value, and the
quantitative basis for this uncertainty relies on the artificial star
experiments. In Appendix A, we further quantitatively explore the
effects of crowding on Cepheids by blue main-sequence stars, and find,
in agreement with the results of Ferrarese {\it et al}., that crowding
effects in the mean are less than 1\%.

\noindent {\it Large-Scale Flows:} Early Cold Dark Matter (CDM) models
of large-scale structure ({\it e.g.,} Turner {\it et al}. 1992)
suggested that sparse sampling of cosmologically small volumes could
give rise to biased values of the Hubble constant (at the 2-4\% level
in samples only extending out to 10,000~km s$^{-1}$). For scales out
to 40,000 km s$^{-1}$, variations were predicted to be only
1-2\%. Over time, the data constraining the local Hubble flow have
continued to increase in sample size, depth and precision. Recent
analyses of a sample of well-measured Type Ia supernovae (Hicken {\it
  et al.} 2009) sampling volumes with velocities extending to over
20,000~km s$^{-1}$ suggest that there are no significant systematic
departures of the Hubble constant from its globally-averaged value;
{\it i.e.,} there is no local void.  This is consistent also with the
analysis of Sandage {\it et al}.  (2010); see also Turnbull {\it et
  al}. (2011) and references therein for a recent discussion of bulk
flows.  At present, we conservatively include a 1-$\sigma$ systematic
error of 1\% for large-scale flows.

\subsection{Summary of Systematic Effects}

We summarize our adopted systematic errors for $H_0$ in Table 1. The
errors are also displayed graphically in Figure 3, and compared with
those from the {\it HST} {\it Key Project}.  The current dominant source of
systematic uncertainty remains the absolute zero point; however, this uncertainty
is a factor of three smaller than for the {\it HST} {\it Key Project} results of a
decade ago. Adding the individual contributions in Table 1 in
quadrature yields an overall systematic uncertainty of $\pm$2.1~km
s$^{-1}$ Mpc$^{-1}$, or an uncertainty of 2.8\% for $H_0$ = 74.3 km
s$^{-1}$ Mpc$^{-1}$, which we adopt as the result for this paper.

\section{Comparison with Riess {\it et al}. (2011)}

In this section, we compare our results with those of Riess {\it et
  al.}~(2011).  In their calibration of the Type Ia supernova distance
scale, they consider three routes to calibrating the Cepheid distance
scale: (1) through the maser galaxy, NGC 4258 (2) through the Milky
Way parallax sample of Benedict {\it et al}. (2007) and (3) through
the LMC. Their second and third paths include an additional allowance
in the uncertainty for possible differences in the zero points of the
photometry transferring from their WFC3 photometric system to the
ground-based photometric calibration of the parallax sample. We can
compare our results most directly with Riess {\it et al}. by comparing
our relative LMC distance moduli (their path 3), and establishing a
3.6 $\mu$m calibration for the Type Ia supernova relative distance
scale. Riess {\it et al}. adopt a distance modulus to the LMC of
18.486 $\pm$ 0.076 based on the Cepheid H-band sample of Persson {\it
  et al.} (2004). Their uncertainty includes  the instrumental zero
points, filter transmission functions, etc).  Adopting this distance
to the LMC, they found $H_0$= 74.4 $\pm$ 2.5~km s$^{-1}$
Mpc$^{-1}$. Combining the zero points from the Milky Way, LMC and NGC
4258, they adopt a final value of $H_0$ = 73.8 $\pm$ 2.3~km s$^{-1}$
Mpc$^{-1}$.

The LMC distance adopted by Riess {\it et al}. (2011) is in very good
agreement with our new distance to the LMC, which is based entirely on
new and completely independent 3.6 $\mu$m Milky Way data from Monson
{\it et al}. (2012) and LMC data from Scowcroft {\it et al}.  (2011,
2012).  The Riess {\it et al}. (2011) H$_0$ calibration makes use of a
larger set of supernova data for which the statistical uncertainties
in the Hubble diagram are decreased to 0.5\%. The excellent agreement
of the new {\it Spitzer} calibration of $H_0$ with that of Riess {\it
  et al.}  provides an independent check on both the value of and the
current systematic uncertainties in $H_0$.


 \section{Comparison of Measurements of the Distance to the LMC}

Benedict et al. (2007) derived a K-band true distance modulus to
the LMC of 18.48 $\pm$ 0.04~mag (their Table 15), based on their Milky
Way calibration applied to the near-infrared data of Persson et
al. (2004).  Riess et al. (2011) review the eclipsing-binary data used
to derive geometric distances to the LMC and, as noted above,  quote an averaged value
of 18.486 $\pm$ 0.065~mag. Applying the Riess et al. (2009) H-band
Milky Way calibration to the Persson et al. (2004) data gives a true
H-band distance modulus to the LMC of 18.49~mag (corrected for E(B-V)
= 0.10~mag).  As pointed out by Benedict et al. (2007) this
tight correspondence of distance moduli for different wavelengths and in
comparison with a geometric distance determination,  suggests again
that any metallicity effect at long wavelengths is small. 

Recently, Walker (2011) has  reviewed the status of determinations of
the distance to the LMC using five independent distance indicators
including Cepheids, RR Lyrae stars, Eclipsing Binaries, Red Variables
and Red Clump stars. These various methods yield an average distance
modulus to the LMC of 18.48 with a full range of 0.1 mag {\it i.e.,}
$\sigma \sim \pm 1.5\%$). Overall, the agreement with the value
obtained in this study is excellent.

\section{Cosmological Implications}

Given a value of $H_{0}$ accurate to $\sim\pm3\%$, what constraints
can we place on cosmological parameters? Of particular interest in
the current era of Dark Energy missions is the dark energy equation
of state, $w_{0}$. As discussed earlier, within the anisotropy
spectrum for cosmic microwave background fluctuations there exist
strong degeneracies between $w_{0}$ and $H_{0}$ ({\it e.g.}, Efstathiou
\& Bond 1999; Hu 2005). An increase in the accuracy of $H_{0}$ therefore
provides a direct means of breaking this degeneracy and improving
the limits on $w_{0}$ from current and future CMB anisotropy experiments
(e.g, WMAP and Planck).

No one experiment can constrain all of the degrees of freedom
describing the current cosmological model. In order to make progress,
we therefore must restrict our parameter space ({\it e.g.}, assuming a
flat universe ($\Omega_{k}=0$) or $w_{0}=$constant).  To open up the
parameter space requires combining different sets of experimental
data, each with their own errors and systematics. Fortunately,
Bayesian inference offers a straightforward way to combine
experimental data, either through the computation of the likelihoods
given each set of data, or by imposing priors. In the case of the CHP,
our constraints on $H_{0}$, being derived from the local distance
scale, are independent of the other cosmological parameters and our
data can therefore be incorporated as a simple Gaussian prior,
centered on $H_{0}=74.3$ ~km s$^{-1}$ Mpc$^{-1}$ with a width
$\sigma=2.1$ ~km s$^{-1}$ Mpc$^{-1}$.

To investigate the constraints on $w_{0}$, we use the Markov Chain
Monte Carlo (MCMC) code COSMOMC developed by Lewis \& Bridle
(2002)\footnote{ \url{http://cosmologist.info/cosmomc}}. To
incorporate our CHP prior, we modified COSMOMC to include the run-time
prior add-on written by Adam Mantz\footnote{
  \url{http://www.slac.stanford.edu/\~amantz/work/cosmomc\_priors/}}.
When incorporating data from Type supernovae (SNe) Ia, we use further
modifications by Alex Conley\footnote{
  \url{http://casa.colorado.edu/~aaconley/cosmomc\_snls/}}.

We begin by combining our result with the WMAP7 cosmic microwave
background anisotropy measurements (Komatsu {\it et al}. 2011). We assume a
flat universe ($\Omega_{k}=0)$ and a dark energy equation of state
that does not evolve with time ($w_{a}=0$). Given these constraints,
the resulting best-fit values for the equation of state parameter is
$w_{0}=-1.09\pm0.10$.  In Figure \ref{fig:WMAP7+BAO}a, we show 1- and
2-$\sigma$ confidence regions in the $H_{0}$/$w_{0}$ plane using both
the CHP constraint on $H_{0}$ (blue curve) and the constraint from F01
(red curve). By way of comparison, the most recent results combining Type Ia supernovae
(SNe Ia), WMAP7, and baryon acoustic oscillations (BAO) under the
same constraints ($\Omega_{k}=0$ and $w_{a}=0$) give $w_{0}=-0.997_{-0.082}^{+0.077}$
(Amanullah et al., 2010). Both results are compatible with a simple
cosmological constant. It should be noted that because the predictive
power of SNe Ia is in the measurement of the curvature of the Hubble
diagram (and not its zero-point), any constraints on dark energy derived
from SNe Ia are insensitive to improvements in the accuracy of $H_{0}$.

We next incorporate  results from baryon acoustic
oscillations (BAO). Unlike SNe Ia, BAO experiments lack a low-redshift
measurement and in a combined analysis of this type, the results are
therefore  constrained by an accurate value of $H_{0}$. Keeping the same
constraints as before ($\Omega_{k}=0$ and $w_{a}=0$) and adding in the
data from Reid et al. (2010), we find $w_{0}=-1.1 \pm 0.10$.  We also
find that the posterior distribution of $H_{0}$ has been shifted down
to $H_0 =72.7 \pm 2.0$ ~km s$^{-1}$ Mpc$^{-1}$. The earlier, broader
prior from F01 would bring the value of $H_{0}$ down even lower to
$H_{0}=66.0_{-4.5}^{+4.1}$ ~km s$^{-1}$ Mpc$^{-1}$ (see Figure
\ref{fig:WMAP7+BAO}b), resulting in a difference between the value of
$H_{0}$ from CMB+BAO alone and that derived from the CHP.  Pushing the
data a little further, we can remove the constraint for  a flat universe. Again,
keeping $w_{a}=0$ and combining WMAP7 and BAO, the constraints on
$w_{0}$ are reduced considerably, resulting in a best-fit value of
$w_{0}=-1.38 \pm 0.24$, which is inconsistent with a
cosmological constant at about 1.5-$\sigma$. The discrepancy between the
prior and posterior values of $H_{0}$ is reduced somewhat, owing to
the increased degrees of freedom in this model. The resulting
constraint on the curvature parameter is $\Omega_{k}=-0.013\pm0.007$,
favoring a flat universe at almost 2-$\sigma$.

We also explore increasing the effective number of relativistic
particles or number of neutrino species, ($N_{eff}$).  We do not
consider limits on neutrino masses here. There is a long literature on
the subject of constraints on neutrino physics (masses and numbers of
particles) from cosmology (see Ma \& Bertschinger 1995; Dolgov 2002;
Lesgourgues and Pastor 2006).  During the radiation era, neutrinos
play a significant role in the cosmological expansion. Neutrinos also
affect the growth of structure, and alter the amplitudes of the peaks
in the cosmic microwave background spectrum, both suppressing and
shifting the positions of the acoustic CMB peaks. The effective number
of neutrino species in the standard model of particle physics is
N${_{eff}}$ = 3.046. The presence of extra relativistic particle species
can lead to measureable effects in the CMB spectrum ({\it e.g.},
Dunkley {\it et al}. 2009; Komatsu {\it et al}. 2009; 2011).

Assuming the extra relativistic particles are massless neutrinos,
their density can be related to the density of photons through
$\rho_{\nu}=0.2271N_{eff}\rho_{\gamma}$ (Komatsu {\it et al}. 2009). This
then modifies the evolution of the Hubble parameter, $H(z)$, by
replacing the standard photon density parameter with
$\Omega_{\gamma}^{\prime}=\Omega_{\gamma}\left(1+0.2271N_{eff}\right)$.
Figure \ref{fig:Neff} illustrates the constraints on the number of
neutrino species, adopting the Komatsu {\it et al}. (2009) model and
combining the CHP H$_0$ value with WMAP7 and BAO data. To begin, we
assume a flat universe. We find $N_{eff}=4.8 \pm 1.0$. In this case
the agreement in the value of $H_{0}$ is improved, but now the
equation of state parameter shifts to a higher value:
$w_{0}=-0.85\pm0.14$.  If we further restrict the model to a pure
cosmological constant ($w_{0}=-1$), the constraints tighten to
$N_{eff}=4.1\pm0.5$, differing by 2-$\sigma$ from the standard value.

In order to investigate the case where both $\Omega_{k}$ and $N_{eff}$
are free to vary, we need to incorporate an additional independent dataset.  We
use the SN Ia data from Sullivan et al. (2011), consisting of the the
SNLS 3-year sample, the SDSSII SN sample, the high-z sample from Riess
et al., (2007), and several low-redshift samples from the literature.
As we mentioned earlier, supernovae have the advantage that they can
be observed at low redshift and therefore do not require $H_{0}$ to
constrain $w_{0}$. Nevertheless, current observations of supernovae
extend only to redshift $z=1.4$ and therefore do not probe the epoch
when radiation was more important. With this increased dataset, we now
relax both restrictions on $\Omega_{k}$ and $N_{eff}$. Figure
\ref{fig:WMAP+BAO+SNLS} shows the combined results of WMAP7, SN Ia,
and BAO. Once again, the red contours show the earlier constraints from F01
and the blue contours show the constraints using the CHP results. The CHP
data do not improve the constraints on $w_{0}$ in this scenario, yet
they still improve the constraints on the number of neutrinos:
$N_{eff}=4.13 \pm 0.67$.  A summary of these cosmological constraints
are included in Table 2.

These MCMC calculations, incorporating our CHP H$_0$ prior and
combining with the WMAP7 data strongly favor a universe with w$_0$
$\sim$ -1. Adding in additional data from BAO and SNe Ia, they are
consistent at the 2-$\sigma$ level with an additional neutrino
species. Other recent studies have obtained results very similar to
those obtained here ({\it e.g}., Dunkley {\it et al} 2010.; Reid {\it
  et al.}  2010; Riess {\it et al}. 2011; Komatsu {\it et al}. 2011;
Mehta {\it et al}.  2012).  Given the number of degrees of freedom,
and real degeneracies that exist amongst the parameters, caution
should be exercised in interpreting 2-$\sigma$ results.  Future data
({\it e.g.,} from Planck) should settle the question of the number of
neutrino species definitively.

\section{Discussion and Conclusions}

As we saw in Figure 3, we have graphically summarized  the
decrease in each of the systematic uncertainties, comparing the {\it HST}
{\it Key Project} and the CHP.  There are four key systematic
improvements to the Cepheid distance scale that have occurred in the
decade separating this study and the {\it HST} {\it Key Project.} 

First, the {\it Spitzer} 3.6~$\mu$m data provide a zero point that is
about an order of magnitude less sensitive to total line-of-sight
extinction than the optical bands used for the {\it Key Project}.
Systematic uncertainties in the reddening corrections (and
uncertainties in the extinction law itself) are virtually eliminated in
moving to the mid-IR. 

Second, longer-wavelength data are theoretically predicted ({\it e.g.,
} Marconi {\it et al}. 2005; Romaniello {\it et al}. 2008) and
empirically demonstrated (Freedman \& Madore 2010; Freedman {\it et
  al}. 2011; Riess {\it et al}. 2011; this paper) to be less sensitive to
metallicity. Moreover, the Milky Way Cepheid sample (now setting the
zero point) has a metallicity more comparable to those of the majority
of the {\it HST} {\it Key Project} spiral galaxies, thereby
eliminating the bulk of the systematic uncertainty involved in
previously using the (lower-metallicity) LMC Cepheids for the
zero-point calibration.

Third, there are now direct parallax measurements for a representative
sample of Milky Way Cepheids to define geometrically the absolute zero point of
the Leavitt relation (Benedict {\it et al}.  2007).  This new {\it
  Spitzer} Galactic trigonometric zero point eliminates the
long-standing dependence on the distance to the LMC. Long-period LMC
Cepheids simply define the slope and width of the 3.6~$\mu$m Leavitt
law. In addition, independent geometric methods for measuring the
distance to the LMC agree with the Cepheid calibration to within 1.5\%
$rms$ in distance, providing an external check on this new
calibration.

 Fourth, and independently of the CHP, new near-infrared Cepheid distances to galaxies
containing Type Ia supernovae (Riess {\it et al}. 2011), as well as
larger samples of more distant supernovae (Hicken {\it et al}. 2009)
have become available.

Based on our analysis of the {\it Spitzer} data available to date,
combined with data from the Hubble {\it Key Project}, we find $H_0$ =
74.3 $\pm$ 0.4 (statistical) $\pm$ 2.1 (systematic) km s$^{-1}$
Mpc$^{-1}$. This value of $H_0$ is in excellent agreement with that of
the {\it Key Project}, as well as that of Riess {\it et al}.
(2011). Combining this result with constraints from WMAP7 alone yields
a value for the dark energy equation of state of $w_0$ = -1.09 $\pm$
0.10. Further combining BAO and SNe Ia data, and relaxing the
restriction on the numbers of neutrino species results in a model with
$w_0$ = -1.08 $\pm$ 0.10 and $N_{eff}$ = 4.13 $\pm$ 0.67. These data
are compatible with, but do not require the presence of an additional
neutrino species.

The dominant source of systematic uncertainty in our new value of
$H_0$ is the zero-point uncertainty in the Cepheid period-luminosity
relation, currently limited by the small numbers of long-period Cepheids
having trigonometric parallaxes.  Nevertheless, this uncertainty is
now more than a factor of three smaller than the zero-point
uncertainty for the {\it Key Project}.

As outlined in F11, as part of the CHP we
have also already observed Cepheids at 3.6~$\mu$m in a sample of nearby
Local Group galaxies and beyond; we are undertaking several
metallicity tests of the Leavitt relation at 3.6~$\mu$m; we are
measuring a Cepheid distance to the maser galaxy, NGC 4258 at
3.6~$\mu$m; and we are calibrating the Tully-Fisher relation at
mid-infrared wavelengths. Future improvements in the Cepheid zero point will
come with the launch of GAIA, and an increased sample of Milky Way
Cepheids and RR Lyrae stars with accurate parallaxes, which are needed
to better define the zero point of the Leavitt relation. 
Having a value of $H_0$ with an externally well-tested and robust total error
budget of  less than 2\% appears feasible within the next decade.

\medskip
\medskip
This work is based in part on observations made with the {\it Spitzer}
Space Telescope, which is operated by the Jet Propulsion Laboratory,
California Institute of Technology under a contract with NASA. Support
for this work was provided by NASA through an award issued by
JPL/Caltech. We thank the staff of the {\it Spitzer} Science Center
for the rapid processing of the data that went into this and other
papers in the series. Computing resources used for this work were made
possible by a grant from the Ahmanson Foundation. This research made
use of the NASA/IPAC Extragalactic Database (NED).

\vfill\eject
\centerline{\bf References \rm}
\vskip 0.1cm
\vskip 0.1cm

\par\noindent  Allende Prieto, C., Lambert, D.~L., \& Asplund, M.\ 2001, \apjl, 556, L63

\par\noindent  Amanullah, R., Lidman, C., Rubin, D., et al.\ 2010, \apj, 716, 712  

\par\noindent Benedict, G. F., McArthur, B.E., Feast, M.W., et al.  2007, AJ, 133, 1810

\par\noindent Dolgov, A. D., 2002, Phys. Rep., 370, 333

\par\noindent Dunkley, J., Komatsu, E., Nolta, M.~R., et al.\ 2009, \apjs, 180, 306

\par\noindent  Efstathiou, G., \& Bond, J.~R.\ 1999, \mnras, 304, 75 

\par\noindent Fazio, G.G., Hora, J. L., Allen, L. E., et al. 2004, ApJS, 154, 10

\par\noindent Ferrarese, L., Silbermann, N.A., Mould, J.R., et
al. 2000, PASP, 112, 177

\par\noindent Freedman, W.L., Madore, B.F., Gibson, B.K., et al. 2001, ApJ, 553, 47 (F01)

\par\noindent Freedman, W.L., \& Madore, B.F. 2010, ARA\&A, 48, 673

\par\noindent Freedman, W.L., \& Madore, B.F. 2011, \apj, 734, 46

\par\noindent Freedman, W.L., Madore, B.F., Scowcroft, V., 
et al. 2011, \aj, 142, 192 (F11)

\par\noindent Hicken, M., Wood-Vasey, W.M., Blondin, S., et al.  2010, \apj, 700, 1097

\par\noindent Hu, W, 2005.  ASP Conf. Series, {\it Observing Dark
  Energy}, 339, 215

\par\noindent Komatsu, E., Dunkley, J., Nolta, M.~R., et al.,  2009, \apjs, 180, 330 

\par\noindent
Komatsu, E., Smith, K.M., Dunkley, J., et al. 2011, ApJS, 192, 18

\par\noindent Lesgourgues, J., \& Pastor, S. 2006, \physrep, 429, 307 

\par\noindent Lewis, A., \& Bridle, S.\ 2002, \prd, 66, 103511

\par\noindent Ma, C.-P., \& Bertschinger, E.\ 1995, \apj, 455, 7 

\par\noindent Marconi, M., Musella, I., \& Fiorentino, G. 2005, ApJ, 632, 590

\par\noindent Mehta, K.~T., Cuesta, A.~J., Xu, X.,  2012, arXiv:1202.0092 

\par\noindent Meixner, M., Gordon, K.~D., Indebetouw, R., et al.\ 2006, \aj, 132, 2268 

\par\noindent Monson, A., Freedman, W.L., Madore, B.F., et al.  2012, \apj, submitted

\par\noindent 
Persson, S.E., Madore, B.F., Krzeminski, W., et al. 2004, AJ, 128, 2239

\par\noindent Reid, B.~A., Percival, W.~J., Eisenstein, D.~J., et al. 2010, MNRAS, 404, 60

\par\noindent 
Rieke, G.H., \& Lebofsky, M.J. 1985, ApJ, 288, 618 

\par\noindent Riess, A.~G., Strolger, L.-G., Casertano, S., et al.\ 2007, \apj, 659, 98 

\par\noindent Riess, A.~G., Macri, L., Casertano, S., et al.\ 2009, \apj, 699, 539 

\par\noindent 
Riess, A.G., Macri, L., Casertano, S., et al.  2011, ApJ, 730, 119

\par\noindent 
Romaniello, M., Primas, F., Mottini, M. et al. 2008, A\&Ap, 488, 731

\par\noindent Sandage, A., Reindl, B., \& Tammann, G.~A.\ 2010, \apj, 714, 1441

\par\noindent 
Scowcroft, V., Freedman, W.L., Madore, B.F.,  et al.  2011, \apj, 743, 76

\par\noindent 
Scowcroft, V., Freedman, W.L., Madore, B.F.,  et al.  2012, \apj, 747, 84 

\par\noindent Stetson, P.~B.\ 1998, \pasp, 110, 1448 

\par\noindent Sullivan, M., Guy, J., Conley, A., et al.\ 2011, \apj, 737, 102 

\par\noindent
Suyu, S. H., Treu,T., Blandford,  R. D.,  et al. 2012, arXiv:1202.4459 

\par\noindent
Turnbull, S.~J., Hudson, M.~J., Feldman, H.~A., et al.\ 2011, \mnras, 2088 

\par\noindent 
Turner, E.L., Cen, R., \& Ostriker, J.P. 1992, AJ, 103, 1427

\par\noindent
Walker, A.R., 2011, Astrophys \& Sp Sci, arXiv:1112.3171 

\par\noindent 
Werner, M.W., Roellig, T. L., Low, F. J.  et al. 2004, ApJS, 154, 1

\par\noindent
Windmark, F., Lindegren, L., \& Hobbs, D.  2011, \aap, 530, A76

\vfill\eject

\vfill\eject
\section{Appendix A: Crowding Errors}

We present here a quantitative discussion of the effects of crowding
on the errors in magnitudes and colors of Cepheids as used to determine
distance in the {\it HST} {\it Key Project} (Freedman {\it et al}. 2001). The Key
Project galaxy fields are dominated by a blue plume of high-mass,
high-luminosity O and B supergiants, which are the longer-lived
progenitors of the Cepheid variables. The blue plume stars are
therefore the most likely objects to be crowding the Cepheids and
contaminating their photometry.

Figure A1 illustrates the effects of the contamination of a Cepheid by
blue main sequence stars. The uncontaminated Cepheid is given a
fiducial magnitude of V = 0.0~mag and a typical Cepheid color of (V-I)
= 1.00~mag. A sequence of progressively brighter main sequence stars,
each having (V-I) = 0.2~mag, was then sequentially added to the light
of the Cepheid. The combined light of the two is plotted as circled
dots progressively making the contaminated Cepheid appear brighter and
bluer. The contaminated Cepheid appears fainter and redder if an uncorrected statistical excess of
blue-plume light was contained in the sky aperture. 

``Fainter and redder'' has the same sense of direction in the
color-magnitude diagram as extinction/reddening.  For the {\it Key
  Project}, reddening was accounted for by producing a Wesenheit
reddening-free magnitude $W =$ V-R$\times$(V-I), where R is the ratio
of total-to-selective absorption such that R = A$_V$/E(V-I), where
A$_V$ is the extinction and E(V-I) is the reddening, and R in this
case has the independently determined value of 2.45. A line of
constant $W$ passing through our sample Cepheid is shown as the solid
line crossing the color-magnitude diagram diagonally from upper left
to lower right. This line also closely tracks the contamination
trajectory defined by the circled dots.  

Quantitatively, for contaminating stars up to 1.8 mag brighter than
the Cepheids, the contaminated Cepheids are only at most 0.06~mag in V
away from the line of constant $W$. In fact the average difference in
$W$ is only 0.03~mag over the range where the V magnitude increases by
a full magnitude due to contamination.  At low levels of contamination
this effect is smaller by a factor of four. For the illustrative
purposes here, we have not modeled the luminosity function of the blue
plume stars. We note, however, that the slope of the luminosity
function is such that contamination by brighter stars is statistically
less likely than for fainter stars. These results are quantitatively
consistent with the artificial star experiments of Ferrarese {\it et
  al}. (2000), who concluded that the effects of crowding in the {\it
  Key Project} fields were less than 0.02 mag or 1\% in
distance. Consequently, as discussed in \S \ref{sec:systematics} and
shown in Figure 3, we have adopted a current uncertainty of 1\% for
this effect.

\begin{table}
\caption{CHP Dominant Systematic Uncertainties in $H_0$}\label{tab1}
\begin{tabular}{@{}lll@{}}
\hline
Source &   Uncertainty & Section / Description  \\
\hline
Absolute Zero Point      &  1.7\% &  \S 3.1 New Milky Way parallaxes + {\it Spitzer} \\
Metallicity Dependence       & 1.4\%  & \S 3.2  Milky Way as reference galaxy \\
WFPC2 Zero Point       &  1\% &   \S 3.3 Ground-based tie-in \\
Crowding       & 1\%  & \S 3.3  Artificial star tests \\
Large-Scale Flows       &  1\% &  \S 3.3 Recent large-scale supernova and galaxy surveys \\
\hline
\end{tabular}
Final Adopted Value: $H_0$ = 74.3 $\pm$ 0.4 (statistical) $\pm$ 2.1 (systematic)~km s$^{-1}$Mpc$^{-1}$\\
Percent Error: \hspace{3.1cm}[$\pm$ 0.5\%] \hspace{1.6cm}[$\pm$ 2.8\%]
\end{table}

\pagebreak

\begin{deluxetable}{lllll}
\tablecolumns{5}
\tablecaption{Constraints on Cosmological Parameters}
\tablehead{\colhead{Dataset/Priors} &
   \colhead{$\Omega_k$} &
   \colhead{$\Omega_m$} &
   \colhead{$w_0$} &
   \colhead{$N_{eff}$}}
\startdata
H$_0$+WMAP7 ($\Omega_k=0,\ N_{eff}=3$) & \nodata & $0.246\pm0.016$ & $-1.09 \pm 0.10$ & \nodata \\
H$_0$+WMAP7+BAO ($\Omega_k=0,\ N_{eff}=3$) & \nodata & $0.263\pm0.015$ & $-1.11\pm0.11$ & \nodata \\
H$_0$+WMAP7+BAO ($N_{eff}=3$) & $-0.013\pm 0.008$ & $0.253\pm 0.016$ & $-1.38\pm 0.24$ & \nodata \\
H$_0$+WMAP7+BAO ($\Omega_k=0$) & \nodata & $0.296\pm 0.027$ & $-0.88\pm 0.15$ & $4.8\pm 1.0$ \\
H$_0$+WMAP7+BAO+SNLS & $-0.007\pm 0.007$ & $0.278\pm 0.018$ & $-1.08\pm 0.10$ & $4.13\pm 0.67$\\

\enddata

Notes: The values quoted are medians of the PPD. The errors are computed by finding the interval over which 68\% of the probability is contained. 
\end{deluxetable}

\pagebreak

\begin{figure*}
    \centering
  \includegraphics[width=12.0cm, angle=0]{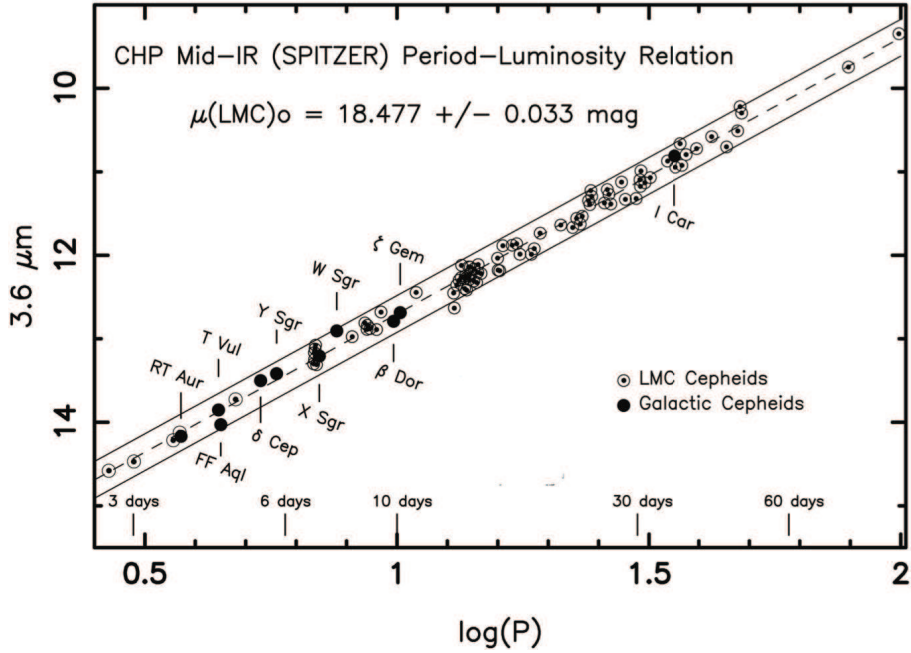}
\caption{The Leavitt law at 3.6~$\mu$m for 80 LMC Cepheids and ten
  Milky Way Cepheids with {\it HST} trigonometric parallaxes.The Milky
  Way data are from Monson {\it et al}. (2012) and the LMC sample is
  from Scowcroft {\it et al}. (2012). The data have been corrected for
  extinction. Small circled points are LMC Cepheids; large filled
  circles, individually named, are Galactic Cepheids with
  trigonometric parallax measurements. The slope of the Leavitt
  relation is set by the LMC sample. Applying this slope to the Milky
  Way sample yields a reddening-corrected distance modulus of
  18.477~mag to the LMC. The five LMC points with periods less than 6
  days are from the sample of Meixner {\it et al}. (2006). They are shown
  for illustration only, and are not included in the fit to determine
  the slope. The dashed slope is defined by the sample of 80 LMC
  stars; the solid lines are 2-$\sigma$ ridge lines. }
\end{figure*}

\begin{figure}
\includegraphics [width=9.0cm, angle=270] {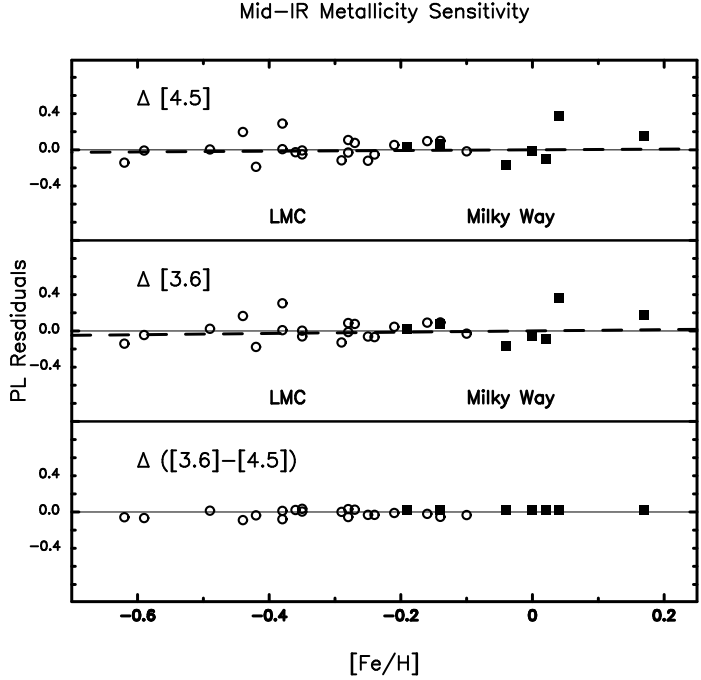}
\caption{Sensitivity of mid-infrared Cepheid magnitudes to
  metallicity. The (lower-metallicity) Large Magellanic Cloud Cepheids
  are shown as open circles; the (higher-metallicity) Milky Way
  Cepheids are plotted as filled squares.  The PL residuals are
  measured in the sense of observed magnitudes minus the mean PL
  relation. The highly correlated nature of the vertical scatter in
  these plots is a reflection of the scatter due to temperature and
  radius variations across the instability strip and to correlated
  back-to-front geometric effects. Neither of these effects are
  expected to correlate with metallicity. No statistically significant
  correlation of the Cepheid magnitudes with atmospheric [Fe/H]
  metallicity can be seen in these plots. The 4.5 $\mu$m residuals are
  plotted in the upper panel; the 3.6 $\mu$m residuals are shown in the
  middle panel. The lower panel shows the remarkably small scatter in
  the color residuals as a function of metallicity where the
  correlated scatter due to instability strip position and
  back-to-front geometry is cancelled. }
\end{figure}

\begin{figure}
\includegraphics [width=9.0cm, angle=270] {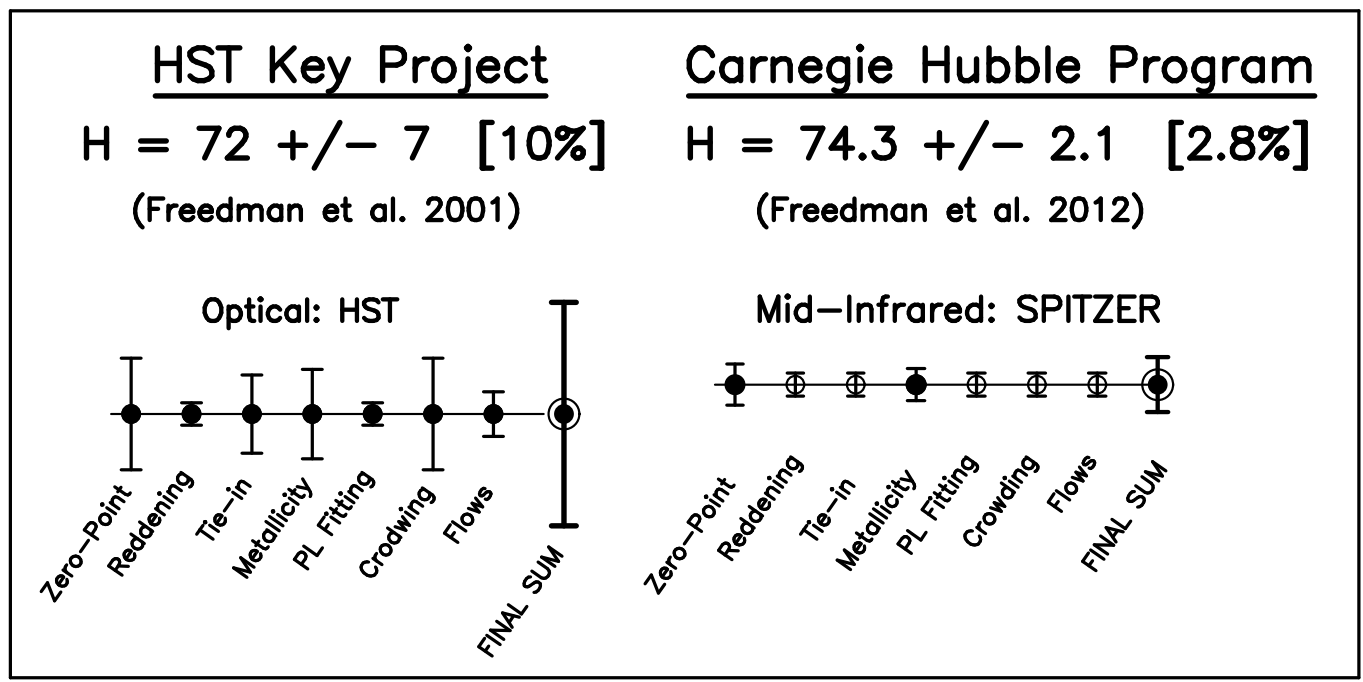}
\caption{Comparison of the seven dominant sources of systematic error
  in determining the Hubble constant. The left-hand panel shows the
  individual errors given for the {\it HST} {\it Key Project} by
  Freedman et al. (2001) (filled circles), followed by the total
  systematic error formed from the quadrature sum of the six preceding
  values (circled dots).  The right-hand portion of the panel shows
  the current errors for the same terms for the CHP determination of
  the Hubble constant. Open circles represent systematic terms that
  have estimated errors less than or equal to 1\%. As discussed in \S
  \ref{sec:systematics}, the sharp drop in the crowding error on the
  right-hand side is based on both the new simulations described in
  Appendix A, as well as the original analysis by Ferrarese {\it et
    al}. 2000. The lower error for the tie-in error results from
  correcting an error in Table 14 of Freedman {\it et al}. (2001),
  which reported the magnitude of a photometric zero-point correction
  and not its uncertainty. The decrease in the zero-point and
  metallicity uncertainties result from the new {\it Spitzer} data.  }
\end{figure}

\begin{figure}
a)\includegraphics[width=3in]{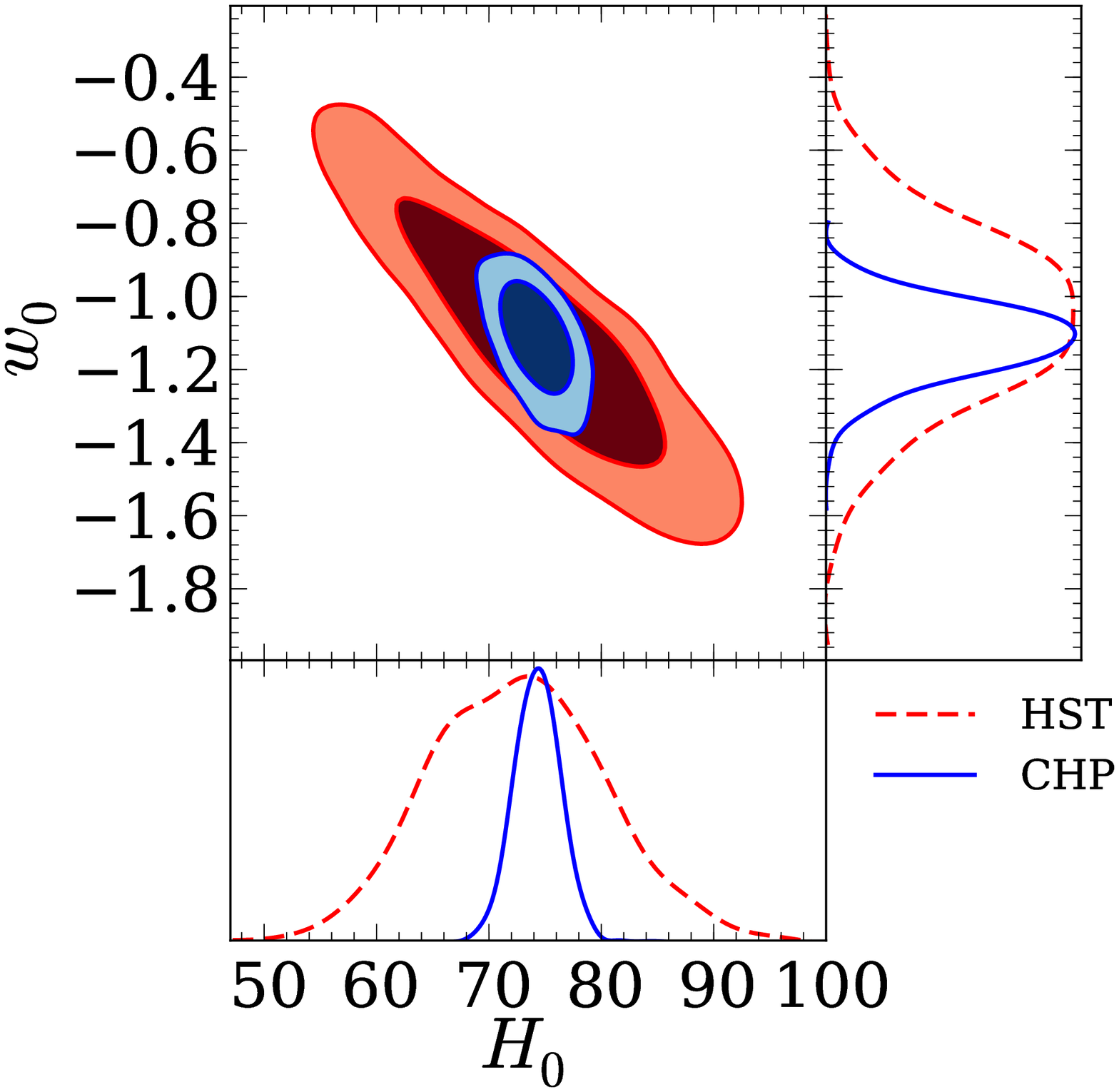}b)\includegraphics[width=3in]{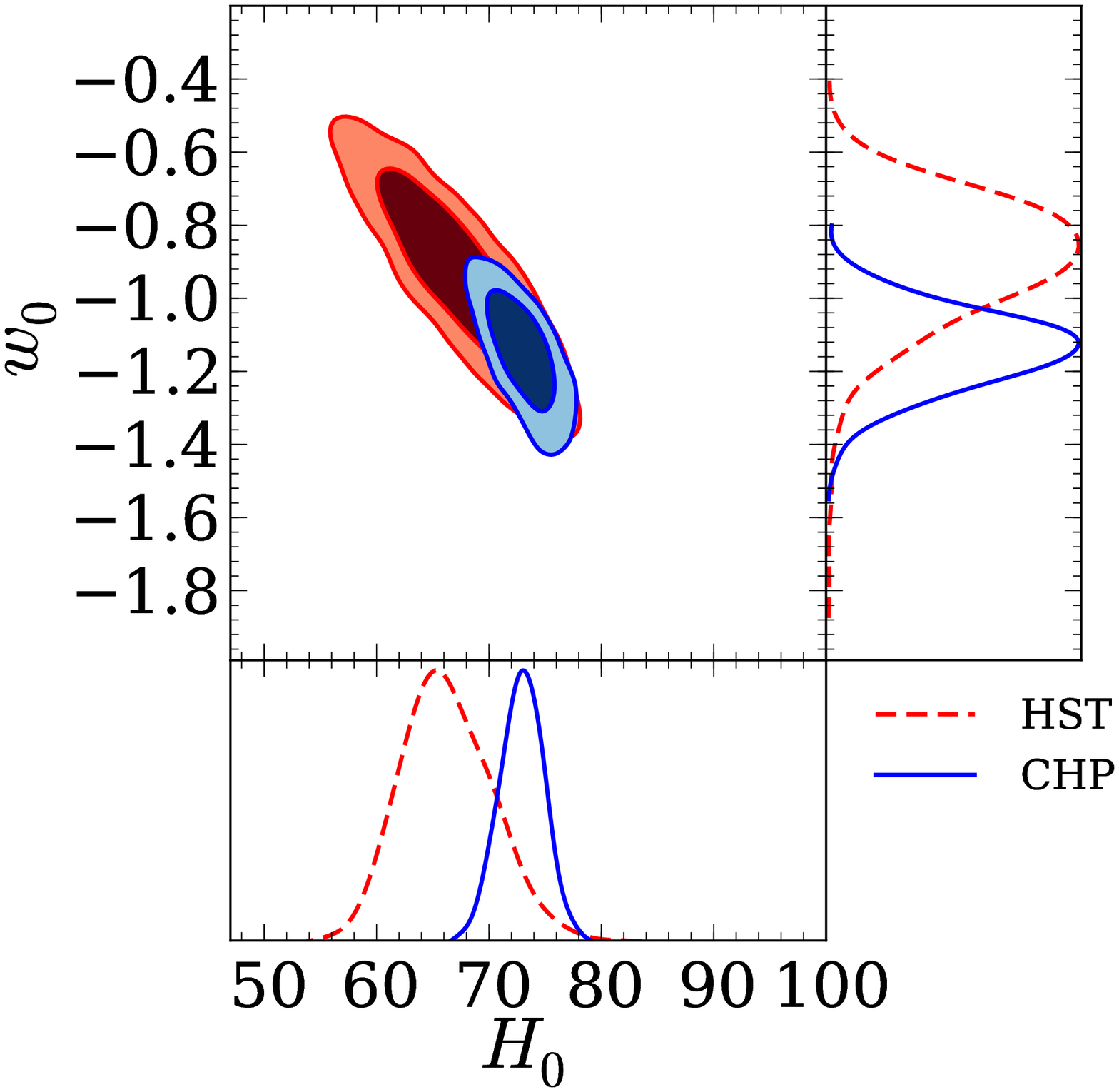}
\caption{2D confidence plots of the equation of state parameter $w_{0}$ and
the Hubble constant $H_{0}$ using (a) the WMAP7 data alone and (b)
WMAP7 and BAO data combined, and assuming $\Omega_{k}=0$, $w_{a}=0$,
and $N_{eff}=3.046$. The red contours show the results using the prior
from F01, while the blue contours show the results using the prior
from this paper (labeled CHP). The right and bottom panels show the
1D marginalized posterior probability distributions (PPD) for $w_{0}$
and $H_{0}$, respectively. The F01 PPD is plotted as red dashed lines,
the CHP PPD is plotted as blue solid lines.\label{fig:WMAP7+BAO}}
\end{figure}
\begin{figure}
\includegraphics[width=6in]{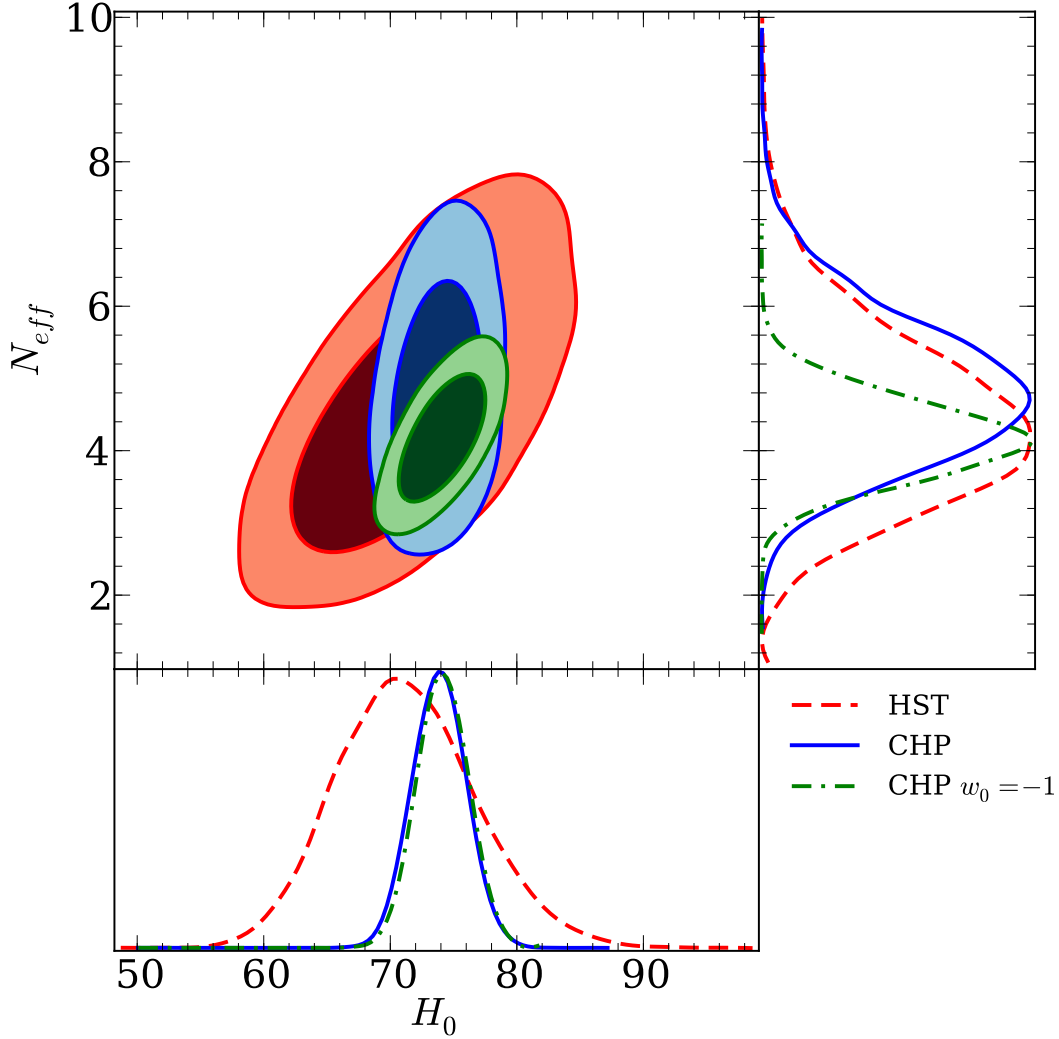}
\caption{2D confidence plot of the effective number of neutrinos
  $N_{eff}$ and the Hubble constant $H_{0}$ using the WMAP7 data and
  BAO data combined and assuming $\Omega_{k}=0$, $w_{a}=0$. The red
  contours show the results using the prior from F01, the blue
  contours show the results using the prior from this paper (labeled
  CHP), and the green contours show the results using the CHP prior
  and assuming $w_{0}=-1$ . The right and bottom panels show the 1D
  marginalized posterior probability distributions (PPD) for $N_{eff}$
  and $H_{0}$, respectively. The F01 ppd is plotted as red dashed
  lines, the CHP PPD is plotted as blue solid lines and the
  CHP+$w_{0}=1$ PPD is plotted as a green dash-dotted
  line.\label{fig:Neff}}

\end{figure}
\begin{figure}
\includegraphics[width=6in]{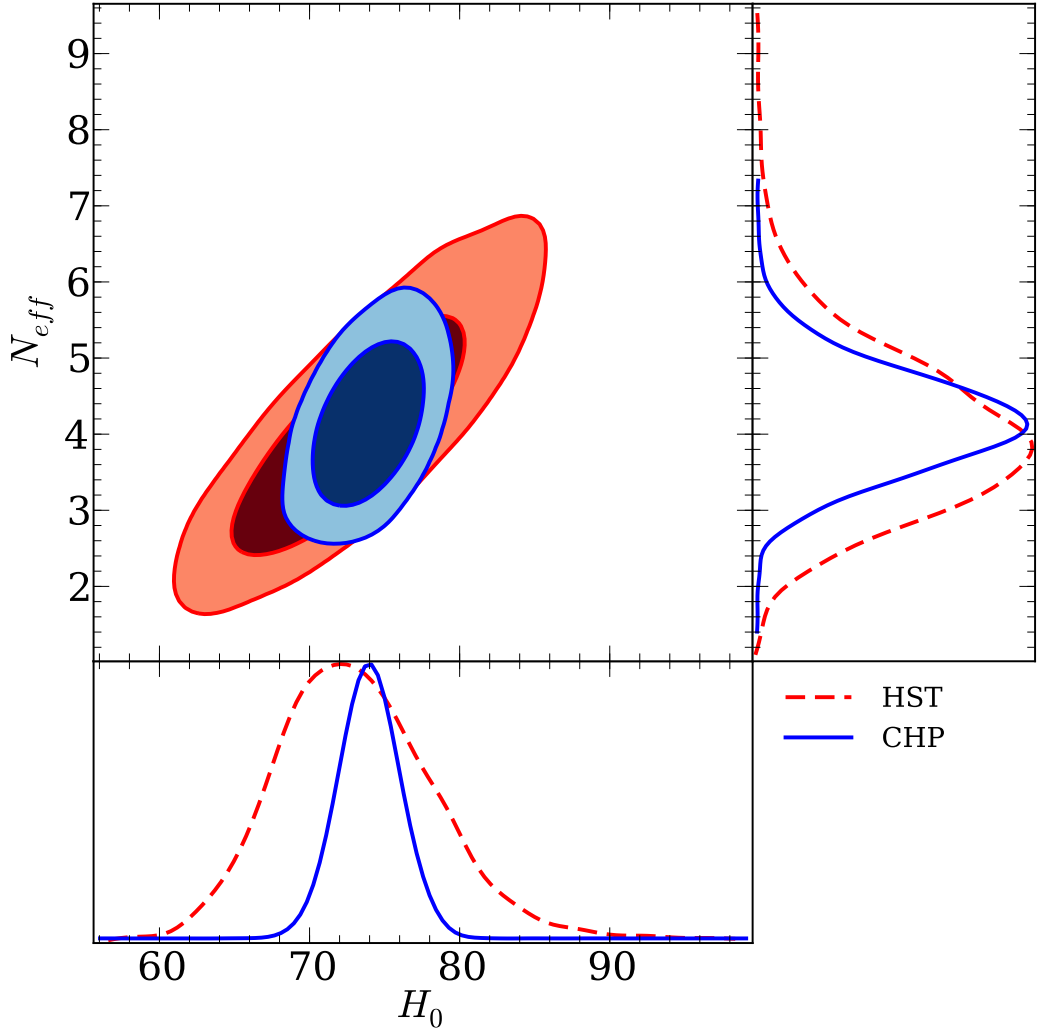}
\caption{2D confidence plot of the effective number of neutrinos $N_{eff}$
and the Hubble constant $H_{0}$ using the WMAP7, BAO, and SN Ia data
combined and assuming $w_{a}=0$. The red contours show the results
using the prior from F01, and the blue contours show the results using
the prior from this paper (labeled CHP). The right and bottom panels
show the 1D marginalized posterior probability distributions (PPD)
for $N_{eff}$ and $H_{0}$, respectively. The F01 PPD is plotted
as red dashed lines, and the CHP PPD is plotted as blue solid lines.
\label{fig:WMAP+BAO+SNLS}}

\end{figure}

\begin{figure}
\includegraphics [width=10cm, angle=-90] {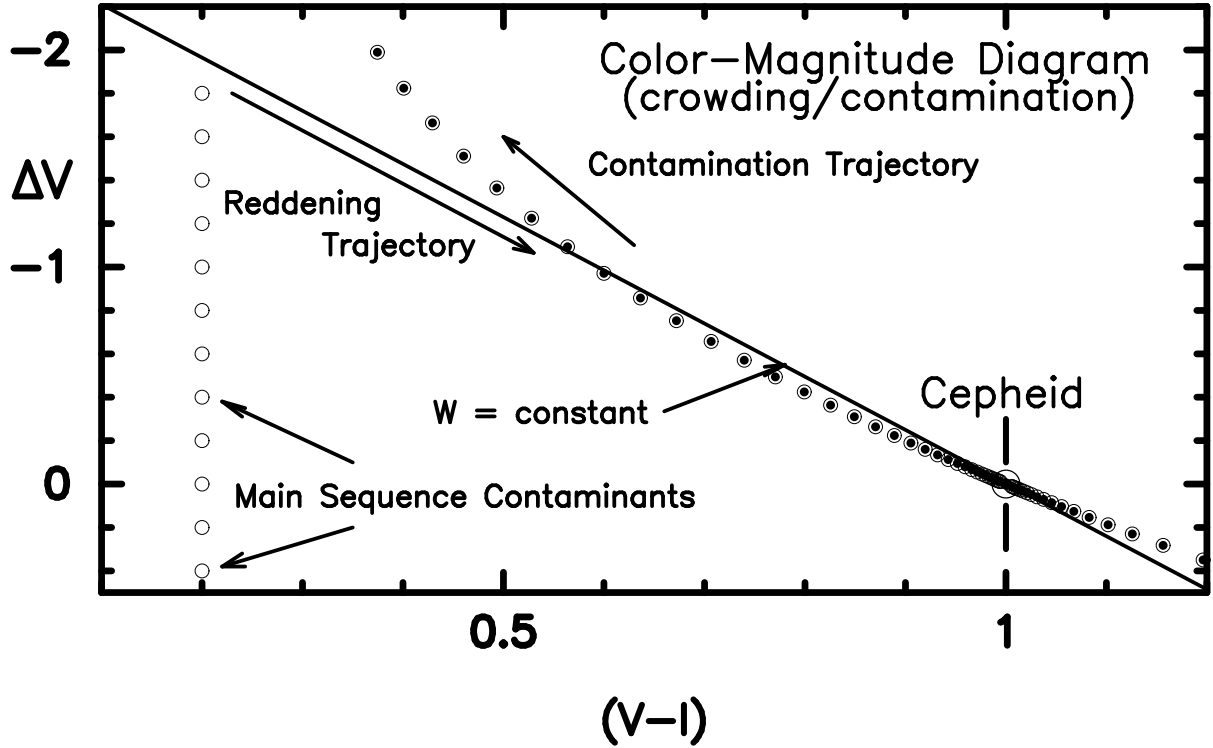}
\caption{(Fig A1): Numerical simulation of the effects of the contamination of a
  Cepheid by an excess of blue main sequence stars under the point spread function
  (artificially brightening the Cepheid) or in the annulus defining
  the sky (thereby artificially dimming the Cepheid). The
  uncontaminated Cepheid, shown by the large circle in the lower right
  of the panel, is set to have fiducial magnitudes and colors of V =
  0.0~mag and (V-I) = 1.00~mag. A sequence of progressively brighter
  main sequence stars (shown as open circles to the left of the panel
  (plotted vertically at (V-I) = 0.2~mag) are sequentially added to
  the light of the Cepheid and the combined light of the two is then
  plotted as a circled dot. The solid line passing through the Cepheid
  and up and to the left is a line of constant $W$, the
  reddening-free magnitude used by the Key Project.}
\end{figure}

\vfill\eject
\end{document}